\documentclass[sigconf]{acmart}

\renewcommand\footnotetextcopyrightpermission[1]{} 
\usepackage{booktabs} 
\usepackage{amsmath}
\usepackage{amssymb}
\usepackage{colortbl}
\usepackage{booktabs}
\usepackage{multirow}
\usepackage{graphicx}
\usepackage{subcaption}
\usepackage{arydshln}
\usepackage{refcount}
\usepackage{balance}
\usepackage[export]{adjustbox}
\graphicspath{ {./images/} }

\begin{document}
\title{Knowledge Query Network for Knowledge Tracing}
\subtitle{How Knowledge Interacts with Skills}

\author{Jinseok Lee}
\affiliation{%
  \institution{Hong Kong University of Science and Technology}
  \city{Hong Kong}
}
\email{benjamin.lee@connect.ust.hk}

\author{Dit-Yan Yeung}
\affiliation{%
  \institution{Hong Kong University of Science and Technology}
  \city{Hong Kong}
}
\email{dyyeung@cse.ust.hk}

\renewcommand{\shortauthors}{J. Lee et al.}

\begin{abstract}
\textit{Knowledge Tracing} (KT) is to trace the knowledge of students as they solve a sequence of problems represented by their related skills. This involves abstract concepts of students' states of knowledge and the interactions between those states and skills. Therefore, a KT model is designed to predict whether students will give correct answers and to describe such abstract concepts. However, existing methods either give relatively low prediction accuracy or fail to explain those concepts intuitively. In this paper, we propose a new model called \textit{Knowledge Query Network (KQN)} to solve these problems. KQN uses neural networks to encode student learning activities into knowledge state and skill vectors, and models the interactions between the two types of vectors with the dot product. Through this, 
we introduce a novel concept called \textit{probabilistic skill similarity} that relates the pairwise cosine and Euclidean distances between skill vectors to the odds ratios of the corresponding skills, which makes KQN interpretable and intuitive.

On four public datasets, we have carried out experiments to show the following: 1. KQN outperforms all the existing KT models based on prediction accuracy. 2. The interaction between the knowledge state and skills can be visualized for interpretation. 3. Based on probabilistic skill similarity, a skill domain can be analyzed with clustering using the distances between the skill vectors of KQN. 4. For different values of the vector space dimensionality, KQN consistently exhibits high prediction accuracy and a strong positive correlation between the distance matrices of the skill vectors.
\end{abstract}

\copyrightyear{2019} 
\acmYear{2019} 
\setcopyright{acmcopyright}
\acmConference[LAK19]{The 9th International Learning Analytics \& Knowledge Conference}{March 4--8, 2019}{Tempe, AZ, USA}
\acmBooktitle{The 9th International Learning Analytics \& Knowledge Conference (LAK19), March 4--8, 2019, Tempe, AZ, USA}
\acmPrice{15.00}
\acmDOI{10.1145/3303772.3303786}
\acmISBN{978-1-4503-6256-6/19/03}

%
%

\ccsdesc[500]{Computing methodologies~Neural networks}
\ccsdesc[500]{Applied computing~E-learning}

\keywords{Knowledge Tracing, Deep Learning, Learning Analytics, Educational Data Mining, Massive Open Online Courses, Intelligent Tutoring Systems, Learner Modeling, Knowledge Modeling, Domain Modeling}

\maketitle
\section{Introduction}
One of the advantages of an Intelligent Tutoring System~\cite{nwana1990intelligent} and massive open online courses~\cite{kaplan2016higher} is that they can potentially benefit from monitoring and tracking student activities in an adaptive learning environment, where learner modeling comes into play. A learner model provides estimates for the students' state, and includes two inter-connected aspects: domain modeling and knowledge modeling~\cite{pelanek2017bayesian}.

A domain model~\cite{pelanek2017bayesian} studies the structure within a domain of problems (For example, it finds out which skill a problem is related to: ``1+2=?" to ``addition of integers" and ``1.3+2.5=?" to ``addition of decimals"). Another task of a domain model is to discover the structure of a skill domain, which can be performed either manually or automatically~\cite{pelanek2017bayesian}. On the other hand, a knowledge model~\cite{anderson1990cognitive}, in an abstract sense, traces students' knowledge while they are solving problems. Knowledge has been described in various forms by the name of knowledge state, which has no universal definition yet. In this paper, the term knowledge state has been used as a state that can describe a student's general level of attainment of skills. Often, domain modeling and knowledge modeling are viewed to be separate; however, we tried to provide approaches for both where problem-solving records of students can be important input features for finding the latent structure of a skill domain.

KT is a research area which analyzes student activities and studies knowledge acquisition, where its main task is to describe a student's knowledge. To elaborate, consider a student who solved a sequence of problems. Then the student's data is given by the temporal sequence of tuples, each of which is consisted of the skill that the problem at each time step is related to and the binary correctness that indicates whether or not the student gave a right answer. By calling such tuple \textit{student response}, the KT problem is formulated as follows: 1. given the student responses up to the time step $t$, describe the student's knowledge state at the current time step $t$, and 2. given the skill at the next time step $t+1$, predict the correctness by modeling the interaction between the student's knowledge state at time $t$ and the skill at time $t+1$, which we will call \textit{knowledge interaction}. Note that the knowledge state refers to the dynamic state of a student accumulated from the student responses while a skill indicates a particular ability that needs to be learned by a student to solve a problem.

Therefore, the quality of a KT model is measured by its ability to describe the knowledge state of a student and its accuracy of predicting correctness. Additionally, since modeling the knowledge interaction is to describe how a student's knowledge state responds to different skills, it is desirable if a KT model can explain the relationship between skills that can be inferred from the knowledge interaction. For example, we can say that ``addition of integers'' is independent of ``subtraction of integers" if a model observes that a student does not learn the latter while learning the former. Similarly, they are dependent if the change in a student's knowledge state of one skill affects the knowledge state of the other. We believe that modeling such skill relationship can lead to further exploration of the latent structure of the skill domain, which is the subject of domain modeling.

However, existing KT models provide limited definitions of either knowledge state, or knowledge interaction, or both. For example, Bayesian Knowledge Tracing (BKT) \cite{anderson1990cognitive} imposes a binary assumption on the knowledge state, which is too restrictive to be intuitive, and Deep Knowledge Tracing (DKT) \cite{piech2015deep} does not give an explanation of knowledge interaction. In this paper, we propose a new neural network KT model called \textit{Knowledge Query Network} (KQN) to generalize the knowledge state and explain the knowledge interaction more descriptively. The central idea is to use the dot product between a knowledge state vector and a skill vector to define the knowledge interaction while leveraging neural networks to encode student responses and skills into vectors of the same $d$ dimensionality. Additionally, we introduce a novel concept called \textit{probabilistic skill similarity} which relates the cosine and Euclidean distances between the skill vectors to the odds ratios for the corresponding skills. Based on those distances, we explore the latent structure of a skill domain with cluster analysis. Lastly, we show that KQN is stable in predicting correctness and learning skill vectors by comparing prediction accuracy and the distance matrices of the skill vectors when the vector space dimensionality is varied.


\section{Related Work}
Item Response Theory (IRT) is a framework for modeling the relationship between problems and correctness \cite{drasgow1990item}. In its simplest form, it uses a logistic regression model by estimating student proficiency and skill difficulty. However, it assumes the proficiency to be constant and does not explain any structure for problems. To overcome those limitations, Bayesian extensions of IRT have been proposed to have a hierarchical structure over items (HIRT) and temporal changes in a student's knowledge state (TIRT) \cite{wilson2016back}. Still, HIRT assumes constant student proficiency while TIRT lacks the ability of domain analysis.

\begin{table}[tbp]
\centering
\begin{subtable}[b]{0.4\linewidth}
\centering
\resizebox{0.6\columnwidth}{!}{\begin{tabular}{rrr} 
\hline
T & SID    & C      \\ 
\hline
1    & 1        & 0            \\
2    & 2        & 0            \\
3    & 1        & 1            \\
\hline
\end{tabular}}
\caption{Original}
\label{input-data-original}
\end{subtable}
\quad
\begin{subtable}[b]{0.5\linewidth}
\centering
\begin{tabular}{rrr}
&&\\
\hline
T & SID & C  \\ 
\hline
1    & 1        & 0            \\
2    & 1        & 1            \\
\hline
\end{tabular}
\qquad
\begin{tabular}{rrr}
&& \\
\hline
T & SID    & C  \\ 
\hline
1    & 2        & 0 \\
\hline
\end{tabular}
\caption{BKT}
\label{input-data-bkt}
\end{subtable}

\begin{subtable}[b]{0.4\linewidth}
\centering
\resizebox{\columnwidth}{!}{\begin{tabular}{rrrr}
&&&\\
\hline
T & SID & NO$_1$ & NO$_2$  \\ 
\hline
1    & 1        & 1   & 0    \\
2    & 2        & 1   & 1    \\
3    & 1        & 2   & 1    \\
\hline
\end{tabular}}
\caption{LFA}
\label{input-data-lfa}
\end{subtable}
\quad
\begin{subtable}[b]{0.5\linewidth}
\centering
\resizebox{0.98\columnwidth}{!}{\begin{tabular}{rrrrrr}
&&&&&\\
\hline
T & SID & S$_1$ & F$_1$ & S$_2$ & F$_2$  \\ 
\hline
1    & 1        & 0  & 1  & 0  & 0   \\
2    & 2        & 0  & 1  & 0  & 1   \\
3    & 1        & 1  & 1  & 0  & 1   \\
\hline
\end{tabular}}
\caption{PFA}
\label{input-data-pfa}
\end{subtable}
\caption{The example student Ben's input data, one original and the others preprocessed for different KT models. In the tables above, feature names are abbreviated as follows: Time to T, Skill ID to SID, Correctness to C, Number of Opportunities to NO, Number of Successes to S, and Number of Failures to F. Note that the original data is used for neural network models.}
\label{input-data}
\end{table}

In BKT, a student's knowledge state is viewed as a set of binary latent variables, one for each skill, with two possible states, \textit{known} and \textit{unknown} \cite{anderson1990cognitive}. Then a set of observable variables, each of which corresponds to correctness per skill, are conditioned on the set of the binary variables. For example, let us say we have a running example of student Ben throughout this paper, who has records as shown in Table \ref{input-data}. Accordingly, there will be two knowledge states and correctness variables for skills 1 and 2, a total of four variables with two independent BKT models with input data as shown in Table \ref{input-data-bkt}. Then the knowledge acquisition in BKT is modeled with a Hidden Markov model (HMM), where knowledge interaction is controlled by a set of interpretable equations. Since BKT lacks the ability to forget and individualize, a number of extensions have been proposed \cite{yudelson2013individualized,khajah2016deep}. However, most importantly, BKT's independence assumption on different skills is considered to be highly constrained and not effective, where the model cannot leverage the whole data. For example, Ben's history of responses on skill 1 cannot tell his responses on skill 2 since there should be two separate models for each skill.

Learning Factors Analysis (LFA) \cite{cen2006learning} models a student's knowledge state as a set of binary variables, one for each skill. Correctness at each time step is predicted with a logistic regression model which has covariates related to students, skills, and a summary statistic of student responses, i.e., the number of past opportunities for each skill. Performance Factors Analysis (PFA) \cite{pavlik2009performance} extends LFA by separating `the number of opportunities per skill' into `the number of correct answers per skill' and `the number of incorrect answers per skill' with the others the same, e.g., in Ben's case, the input data are preprocessed as shown in Table \ref{input-data-lfa} and Table \ref{input-data-pfa}. 

Since an estimate for correctness is explained with student covariates, and skill-specific covariates without variable interactions in LFA and PFA, the two models do not describe how a student's knowledge state with respect to one skill is affected by that with respect to another skill; instead, a student parameter, which is also called student proficiency, is the only factor that relates the knowledge state for different skills. Moreover, a skill is explained by the regression coefficients for the skill-specific covariates from which we cannot tell the structure of a skill domain directly.

As the first neural network KT model, DKT \cite{piech2015deep}, given the student's responses as input, encodes a student's knowledge state as a summarized vector calculated from a recurrent neural network (RNN), which is a renowned neural network technique for modeling temporal data. However, DKT does not define the knowledge interaction directly. In detail, a student response at each time step is formed as a tuple $(q_t, a_t)$, where $q_t$ and $a_t$ refer to the problem ID and correctness, respectively. For example, Ben's original data in Table \ref{input-data-original} is expressed as $(1, 0)$ at $t=1$. The input is then passed to an RNN layer, where Long Short-Term Memory (LSTM) \cite{hochreiter1997long} was used in the original paper \cite{piech2015deep}, and its output hidden state is passed to a logistic function after an affine transformation, i.e., $\mathbf{y}_t = \sigma(\mathbf{W} \cdot \mathbf{h}_t + \mathbf{b})$, where $\mathbf{h}_t$ is the output hidden state of an LSTM layer and $\sigma$ is an element-wise logistic function. Finally, the $k$-th element of $\mathbf{y}_t$ is used to predict correctness at the next time step given that the next problem ID is $k$. Despite its superior prediction performance over the existing classical methods, DKT has been criticized by other papers \cite{wilson2016back,khajah2016deep} for its lack of practicality in educational applications. This is because the output hidden state $\mathbf{h}_t$ is inherently hard to be interpreted as the knowledge state, and the model does not give insights into the knowledge interaction.

To make a neural network KT model more interpretable, Dynamic Key-Value Memory Networks (DKVMN) \cite{zhang2017dynamic} have been introduced by extending memory-augmented neural networks (MANNs) \cite{graves2014neural,DBLP:journals/corr/WestonCB14}. Like DKT, DKVMN uses the original data as input. DKVMN accumulates temporal information from student responses into a dynamic matrix, or the \textit{value memory}, while embedding skills with a static matrix, or the \textit{key memory}. DKVMN defines the interaction between value vectors and key vectors using attention weights calculated with cosine similarity, and predicts correctness by passing a concatenation of a weighted sum of value vectors and an embedded skill vector to a multilayer perceptron (MLP). Even though the prediction accuracy of DKVMN has proven to be higher than that of DKT, the use of an MLP for the output of the model still makes it hard to explain the knowledge interaction.

\section{Our Proposed Model}
\subsection{Motivation}
To generalize the knowledge state while describing the knowledge interaction intuitively, we suggest a model that projects a student's knowledge and skills into the same vector space of embedding dimensionality $d$. An important constraint is to contain the skill vectors on the $d$-dimensional positive orthant unit sphere, i.e., they have unit-length and positive coordinates. The logit of a probability estimate for correctness is given by the dot product between the current knowledge state vector and the skill vector of the next problem. This is only possible because both knowledge state and skill vectors lie in the same vector space.

\begin{figure}[ht]
\includegraphics[width=0.7\columnwidth]{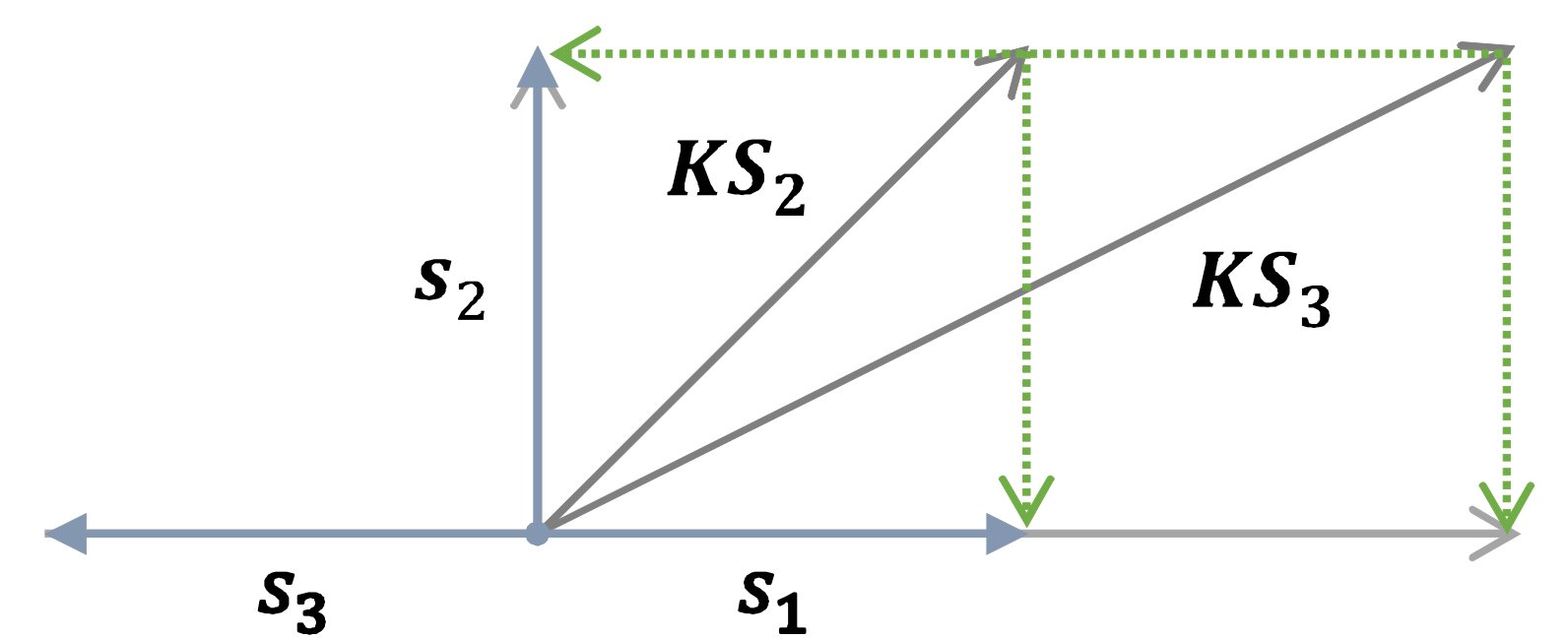}
\centering
\caption{Illustration of skill vectors and Ben's knowledge state vectors at $t=2$ and $t=3$.}
\label{example-illustration}
\end{figure}

Now, we illustrate why skill vectors are set to unit-length and constrained to a positive orthant: the former makes the logit only dependent on the direction of the related skill vector while the latter assures that learning on one skill does not decrease learning on another. For example, in a 2-D vector space, suppose that Ben has knowledge state $\mathbf{KS}_2 = (1, 1)^T$ at $t=2$ after two responses while there are three skill vectors, $\mathbf{s}_1 = (1, 0)^T$ and $\mathbf{s}_2 = (0, 1)^T$ for the skills 1 and 2, and $\mathbf{s}_3 = (-1, 0)^T$ for a third ``imaginary'' skill as shown in Figure \ref{example-illustration}. At $t=3$, his knowledge state may change to $\mathbf{KS}_3 = (2, 1)^T$ as he answers correctly for skill 1. Then the logit with respect to $\mathbf{s}_1$ increases from $\mathbf{KS}_2 \cdot \mathbf{s}_1 = 1$ to $\mathbf{KS}_3 \cdot \mathbf{s}_1 = 2$ while that with respect to skill 2 remains the same as $\mathbf{KS}_2 \cdot \mathbf{s}_2 = \mathbf{KS}_3 \cdot \mathbf{s}_2 = 1$. However, the logit with respect to skill 3 would decrease from $\mathbf{KS}_2 \cdot \mathbf{s}_3 = -1$ to $\mathbf{KS}_3 \cdot \mathbf{s}_3 = -2$, which would then decrease the probability estimate for the correctness of skill 3. This is counter-intuitive since the datasets we are dealing with have a set of skills within the same area, e.g., mathematics.

\subsection{Objective}
Let the skill of a problem and the correctness at time $t$ be $e_t \in \{1, \cdots, N\}$ and $c_t \in \{0, 1\}$, respectively. The correctness $c_{t+1}$ at each time step $t = 1, \cdots, T-1$ is viewed as a Bernoulli variable given the history of responses and the skill at time step $t+1$. Then the objective of a KT model is to find out the parameter of the Bernoulli distribution at each time step as follows:

\begin{align*}
p_{t+1} &= P(c_{t+1} = 1 \text{ } | \text{ } e_{1:t+1}, c_{1:t}), \\
c_{t+1} &\sim \mathit{Bernoulli}(p_{t+1}).
\end{align*}

\subsection{Architecture Overview}
\begin{figure}[tbp]
\includegraphics[width=\linewidth]{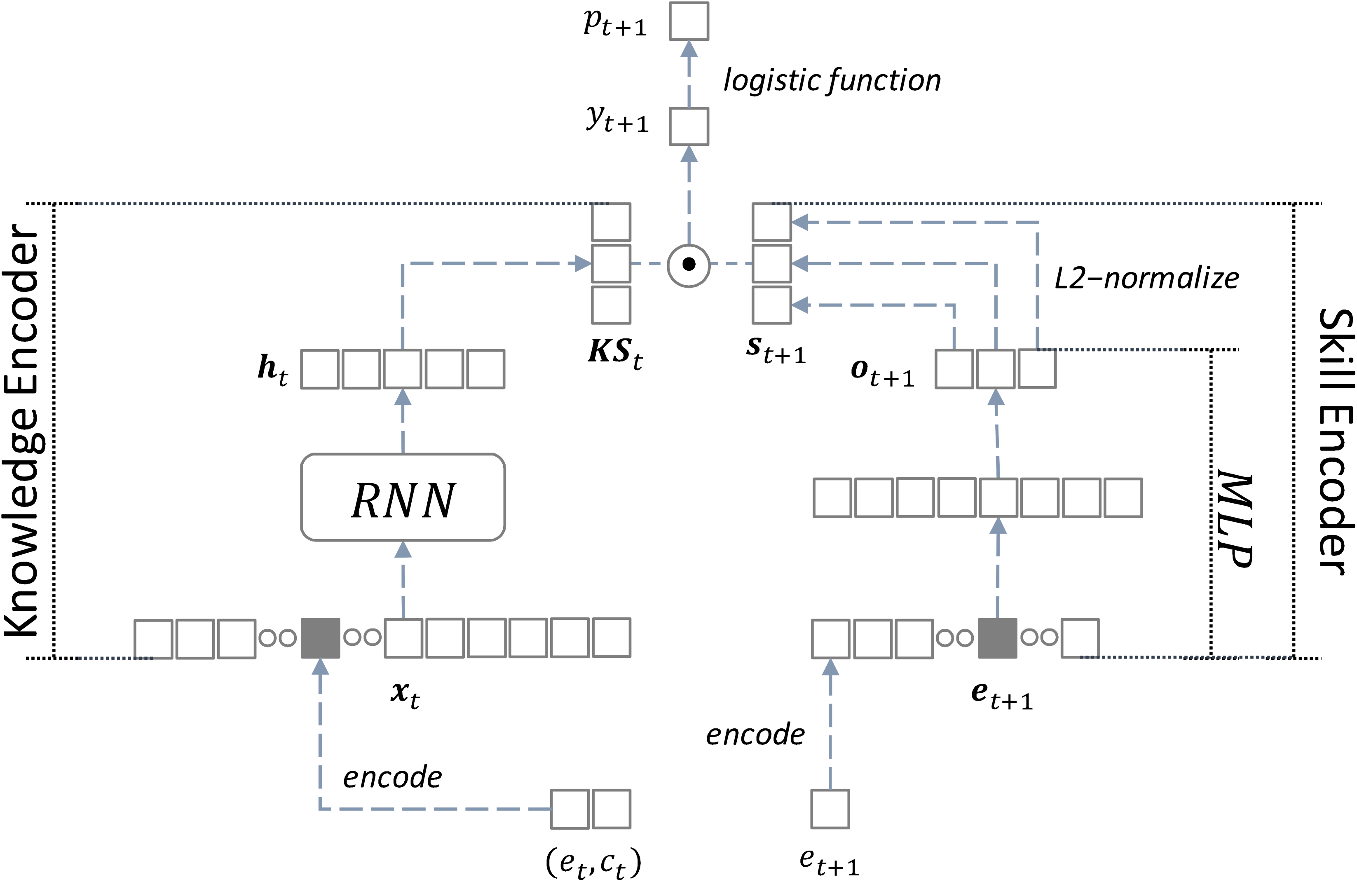}
\centering
\caption{KQN architecture drawn at time $t$.}
\label{model-diagram}
\end{figure}

KQN consists of three components: knowledge encoder, skill encoder, and knowledge state query. The knowledge encoder converts the temporal information from student responses into a knowledge state vector while the skill encoder embeds a skill into a skill vector. The two vectors are then passed to the knowledge state query to provide the prediction for correctness given the current knowledge state and the provided skill. The network architecture of KQN is shown in Figure \ref{model-diagram}.

\subsection{Inputs}
The model takes two inputs: a student response at the current time step and a skill at the next time step. Each of the student responses is one-hot encoded and given as input $\mathbf{x}_t$ to an RNN layer as the following:
\begin{align*} 
&\mathbf{x}_t \in \{0, 1\}^{2N}, \\
&\mathbf{x}_t^k = 1 \text{ if the answer is wrong}, \\
&\mathbf{x}_t^{k+N} = 1 \text{ if the answer is correct}, \\
\end{align*}
where $N$ is the number of skills, and $k$ is the skill at time step $t$. Similarly, the skill $k'$ at time $t+1$ is one-hot encoded to $\mathbf{e}_{t+1}$, where the $k'$-th element is 1 and the other elements are 0's. In Ben's case, his response at $t=1$ and the skill for the problem at $t=2$ are encoded to $\mathbf{x}_1 = (1, 0, 0, 0)^T$ and $\mathbf{e}_2 = (0, 1)^T$, respectively.

\subsection{Knowledge State Query}
Let the knowledge state vector $\mathbf{KS}_t$ and the embedded skill vector $\mathbf{s}_{t+1}$, both $d$-dimensional, be the two vectors encoded by the knowledge encoder and the skill encoder, respectively. Then the knowledge interaction is defined by the inner product of the two vectors. The logit $y_{t+1}$ and the corresponding prediction probability $p_{t+1}$ are calculated as follows:
\begin{align*}
&y_{t+1} = \mathbf{KS}_t \cdot \mathbf{s}_{t+1}, \\
&p_{t+1} = \sigma(y_{t+1}),
\end{align*}
where $\sigma(u) = \frac{1}{1 + \exp(-u)}$ is a logistic function and $\cdot$ refers to the inner product. In this way, knowledge interaction is well-defined for the following reasons:

\begin{itemize}
\item {
If two skills are independent, their corresponding vectors are orthogonal to each other. Accordingly, an increase or a decrease in the logit with respect to one vector does not affect the logit with respect to the other vector.
}
\item {
If two skills are similar from the probabilistic perspective, then an increase in a logit with respect to one vector would lead to an increase in the logit with respect to the other vector, and vice versa.
}
\end{itemize}

Note that from the definition of the knowledge interaction above, it is implied that there can be at most $d$ mutually independent skills. For different values of $d$, whether or not KQN learns the pairwise relationships between skills represented by pairwise distances was tested in experiments and shown in later sections.

As a result, KQN is approximating the parameter of the Bernoulli distribution at each time step as follows:
\begin{align*}
P(c_{t+1} = 1 \text{ } | \text{ } e_{1:t+1}, c_{1:t})
&= P(c_{t+1} = 1 \text{ } | \text{ } \mathbf{x}_{1:t}, \mathbf{e}_{t+1}) \\
&\approx \sigma(y_{t+1}) = \sigma\bigl(\mathbf{KS}_t \cdot \mathbf{s}_{t+1} \bigr).
\end{align*}

\subsection{Knowledge State Encoder}
Given input $\mathbf{x}_t$, the knowledge state encoder produces a knowledge state vector $\mathbf{KS}_t$ with the following equations:
\begin{align*}
&\mathbf{h}_t = \mathit{RNN}(\mathbf{x}_t), \\
&\mathbf{KS}_t = \mathbf{W}_{h,KS} \cdot \mathbf{h}_t + \mathbf{b}_{h, KS},
\end{align*}
where $\mathbf{W}_{h, KS} \in \mathbb{R}^{d \times H_{\mathit{RNN}}}$, $\mathbf{b}_{h, KS} \in \mathbb{R}^d$, $\mathit{RNN}$ is an RNN layer, and $H_{\mathit{RNN}}$ is the state size of $\mathit{RNN}$. In KQN, LSTM \cite{hochreiter1997long} and Gated Recurrent Units (GRU) \cite{cho2014learning} have been tested as RNN variants. Also, to avoid overfitting, dropout regularization \cite{srivastava2014dropout,zaremba2014recurrent} has been used for the RNN output layer as was used for DKT in a previous work \cite{xiong2016going}.

\subsection{Skill Encoder}
The skill encoder embeds input $\mathbf{e}_{t+1}$ to $\mathbf{s}_{t+1}$ with an MLP as follows:
\begin{align*}
&\mathbf{o}_{t+1} = \mathit{ReLU}(\mathbf{W}_1 \cdot (\mathit{ReLU}(\mathbf{W}_0 \cdot \mathbf{e}_{t+1} + \mathbf{b}_0)) + \mathbf{b}_1), \\
&\mathbf{s}_{t+1} = \mathit{L2\text{-}normalize}(\mathbf{o}_{t+1}), \\
&\mathbf{s}_{t+1} \in \mathbf{U}^d = \{\mathbf{v} \in \mathbb{R}^d \text{ : } ||\mathbf{v}|| = 1, \mathbf{v}^i \geq 0 \text{ for } i = 1,\cdots,d\},
\end{align*}
where $\mathbf{W}_0 \in \mathbb{R}^{H_{\mathit{MLP}} \times N}$, $\mathbf{W}_1 \in \mathbb{R}^{d \times H_{\mathit{MLP}}}$, $\mathbf{b}_0 \in \mathbb{R}^{H_{\mathit{MLP}}}$, $\mathbf{b}_1 \in \mathbb{R}^d$, and $\mathit{ReLU}$ is an element-wise ReLU activation with $\mathit{ReLU}(u) = \max(0, u)$ \cite{nair2010rectified}. Note that $\mathbf{s}_{t+1}$ is now constrained to the $d$-dimensional positive orthant unit sphere, which we will call $\mathbf{U}^d$ for the rest of this paper for notational convenience.

\subsection{Optimization}
At each time step, the cross-entropy error given the probability estimate and the target correctness is calculated, and the error terms for $t = 1, \cdots, T-1$ are summed to give the total error.
\begin{align*}
&E(\boldsymbol\theta_{model} \text{ } | \text { } c_{t+1}, p_{t+1}) \\
&= -\biggl[c_{t+1} \log p_{t+1} + (1-c_{t+1}) \log (1-p_{t+1})\biggr], \\
&E_{total}(\boldsymbol\theta_{model} \text{ } | \text{ } c_{2:t+1}, p_{2:t+1}) \\
&= \sum_{t=1}^{T-1} E(\boldsymbol\theta_{model} \text{ } | \text{ } c_{t+1}, p_{t+1}).
\end{align*}

The gradients of the total error with respect to the model parameters $\boldsymbol\theta_{model}$ have been computed with back-propagation to be used by an optimization method.

\section{Probabilistic Skill Similarity}
Based on the architecture of KQN, we hereby introduce a novel concept called \textit{probabilistic skill similarity} to measure the distance between skills from the probabilistic perspective.

\subsection{Distance Measures for Skill Vectors}
For any two skill vectors $\mathbf{s}_1, \mathbf{s}_2$ learned from KQN, the cosine distance differs from the squared Euclidean distance by only a factor of 2 since they are constrained to $\mathbf{U}^d$ as follows:

\begin{align*}
&d_{Euclidean}(\mathbf{s}_1, \mathbf{s}_2)^2 
= ||\mathbf{s}_1 - \mathbf{s}_2||^2 
= 2 (1 - \mathbf{s}_1 \cdot \mathbf{s}_2) \\
&= 2 d_{cosine}(\mathbf{s}_1, \mathbf{s}_2), \text{ } \forall \mathbf{s}_1, \mathbf{s}_2 \in \mathbf{U}^d.
\end{align*}

\subsection{Distances and Odds Ratios}
Next, we show how a pairwise distance between two skill vectors is related to the logarithm of their odds ratio. Given a knowledge state vector $\mathbf{KS} \in \mathbb{R}^d$ and a skill vector $\mathbf{s} \in \mathbf{U}^d$, the probability estimate $p$ for correctness and the corresponding odds $o$ are calculated as follows:

\begin{align*}
p = P(c = 1 \text{ } | \text{ } \mathbf{KS}, \mathbf{s}),\; o = \frac{p}{1-p}.
\end{align*}

Then for any two skill vectors $\mathbf{s}_1, \mathbf{s}_2 \in \mathbf{U}^d$, the logarithm of the odds ratio is characterized by their distance as follows:

\begin{align*}
\biggl(\log \frac{o_1}{o_2}\biggr)^2 &= \bigl(\log o_1 - \log o_2\bigr)^2 \\
&= (y_1 - y_2)^2 \\
&= (\mathbf{KS} \cdot \mathbf{s}_1 - \mathbf{KS} \cdot \mathbf{s}_2)^2 \\
&= \bigl(\mathbf{KS} \cdot (\mathbf{s}_1 - \mathbf{s}_2)\bigr)^2 \\
&= \bigl(\mathbf{KS} \cdot \boldsymbol\Delta_{1,2}\bigr)^2 \times ||\mathbf{s}_1 - \mathbf{s}_2||^2 \\
&= \bigl(\mathbf{KS} \cdot \boldsymbol\Delta_{1,2}\bigr)^2 \times d_{Euclidean}(\mathbf{s}_1, \mathbf{s}_2)^2 \\
&= \bigl(\mathbf{KS} \cdot \boldsymbol\Delta_{1,2}\bigr)^2 \times 2d_{cosine}(\mathbf{s}_1, \mathbf{s}_2),
\end{align*}
where $\boldsymbol\Delta_{1,2} = \frac{\mathbf{s}_1 - \mathbf{s}_2}{||\mathbf{s}_1 - \mathbf{s}_2||}$. Therefore, we say that two skills are \textit{probabilistically similar} if they are `close' enough based on the distance between their corresponding vectors.

\section{Experiments}
KQN has been tested for four tasks: \textit{correctness prediction}, \textit{knowledge interaction visualization}, \textit{skill domain analysis}, and \textit{the sensitivity analysis of the dimensionality of the vector space}. 
For correctness prediction, the performance of KQN was compared to that of other models on four public datasets: one synthetic and three real-world ones which are available online. Then for a sample student, knowledge interaction was visualized with a heat map to demonstrate the knowledge state query with respect to different skills. Next, the skill domain was explored with clustering based on skill distances. Finally, pairwise distances of the skill vectors in one dimensionality were compared to those in other dimensionalities to conduct the sensitivity analysis of the vector embedding dimensionality.

\subsection{Datasets}
The following four datasets have been used to evaluate models: ASSISTments 2009-2010, ASSISTments 2015, OLI Engineering Statics 2011, and Synthetic-5. To make a fair comparison of the correctness prediction task, we used the ones provided by the DKVMN source code available online\footnote{\label{dkvmn-data}https://github.com/jennyzhang0215/DKVMN}. The statistics of the datasets are shown in Table \ref{data-stats}.

\subsubsection{ASSISTments 2009-2010}
It is a dataset\footnote{https://sites.google.com/site/assistmentsdata/home/assistment-2009-2010-data/skill-builder-data-2009-2010} collected by the ASSISTments online tutoring systems \cite{feng2009addressing}. It was gathered from skill builder problem sets, where students work on the problems to achieve mastery, a certain level of performance, working on similar questions. During the preprocessing, those records without skill names have been discarded. After a problem with duplicate records had been reported by a paper \cite{xiong2016going}, the dataset has since been corrected by the ASSISTments system. Therefore, those results reported by a number of previous papers are not compared in this paper.

\subsubsection{ASSISTments 2015}
Compared to ASSISTments 2009-2010 which has 110 distinct skill tags, this dataset\footnote{https://sites.google.com/site/assistmentsdata/home/2015-assistments-skill-builder-data} contains 100 distinct ones with more than twice the number of student responses. Data records with invalid \textit{correct} values that are not in $\{0, 1\}$ have been removed.

\subsubsection{OLI Engineering Statics 2011}
This dataset\footnote{https://pslcdatashop.web.cmu.edu/DatasetInfo?datasetId=507} was gathered from a college level statics course in Fall 2011 \cite{koedinger2010data}. The concatenation of a problem name and a step name has been labeled as a skill tag. Note that the number of skills is much larger than those of other datasets.

\subsubsection{Synthetic-5}
It is a dataset\footnote{https://github.com/chrispiech/DeepKnowledgeTracing} originally generated by the authors of the DKT paper \cite{piech2015deep}. Each student response was generated using skill difficulty, student proficiency, and the probability of a random guess set to a constant based on IRT \cite{drasgow1990item}. The dataset consists of a number of sub-datasets, and those with five concepts from version 0 to version 19 have been used, i.e., a total of 20 sub-datasets from the original dataset were used.

\begin{table}[ht]
\begin{tabular}{@{}lrrrr@{}}
\toprule
\multicolumn{1}{c}{Dataset}          & \multicolumn{1}{c}{Students} & \multicolumn{1}{c}{Skills} & \multicolumn{1}{c}{Size} & \multicolumn{1}{c}{Max Steps} \\ \midrule
ASSIST2009 & 4,151     & 110    & 325,637       & 1,261           \\
ASSIST2015 & 19,840    & 100    & 683,801       & 618            \\
Statics2011     & 333      & 1,223   & 189,297       & 1,181           \\
Synthetic-5      & 4,000     & 50     & 200,000       & 50             \\ \bottomrule
\end{tabular}
\caption{Statistics for all the datasets. Names of the datasets have been abbreviated. `Size' and `Max Steps' refer to the total number of student responses and the maximum number of time steps, respectively.}
\label{data-stats}
\end{table}

\begin{table*}[tbp]
\centering
\begin{tabular}{@{}lcccccc@{}}
\toprule
\multicolumn{1}{c}{\multirow{2}{*}{Dataset}} & \multicolumn{6}{c}{Test AUC (\%)} \\ \cmidrule(l){2-7} 
           & IRT+  & BKT+ & DKVMN     & DKT       & DKT+KQN & KQN  \\ \midrule
ASSIST2009 & 77.40 & -    & 81.57$\pm$0.1 & 80.53$\pm$0.2 & 82.05$\pm$0.04 & \textbf{82.32$\pm$0.05} \\ \midrule
ASSIST2015 & -     & -    & 72.68$\pm$0.1 & 72.52$\pm$0.1 & \textbf{73.41$\pm$0.02} & 73.40$\pm$0.02 \\ \midrule
Statics2011& -     & 75   & 82.84$\pm$0.1 & 80.20$\pm$0.2 & 80.27$\pm$0.22 & \textbf{83.20$\pm$0.05} \\ \midrule
Synthetic-5& -     & 80   & 82.73$\pm$0.1 & 80.34$\pm$0.1 & 82.58$\pm$0.01 & \textbf{82.81$\pm$0.01} \\ \bottomrule
\end{tabular}
\caption{Accuracy of different KT models were compared based on test AUCs (\%) for the correctness prediction task. IRT, BKT, and their variants were used as representatives of non-neural-network KT models. DKVMN and DKT were compared to KQN as baselines of neural-network models while DKVMN has been the previous state-of-the-art neural network KT model. Note that DKT+KQN refers to DKT with the embedded skill vectors learned from KQN.}
\label{prediction-performance}
\end{table*}

\subsection{Setup and Implementation}
All the program codes for the implemented KQN and DKT were written in TensorFlow 1.5\footnote{https://www.tensorflow.org/versions/r!1.5/}. For the data splits of each dataset, we used the same ones\footnotemark[\getrefnumber{dkvmn-data}] used by DKVMN for a fair comparison of prediction accuracy.

\subsubsection{Correctness Prediction}
All the sequences of student responses were preserved in their original length without truncation. Each dataset except Synthetic-5 has been split into training, validation, and test sets with 8:2 and 7:3 for training to validation and (training+validation) to test ratios, respectively. For Synthetic-5, the corresponding ratios of 8:2 and 5:5 have been set. Hyperparameters have been grid-searched with holdout validation with early-stopping. Note that no early-stopping was used in the testing phase. The number of epochs was set to 50 and 200 during the validation and testing phases, respectively. During the testing phase, KQN was run for five times, and the mean and standard deviation of the performance metric results have been reported.

The hyperparameters of KQN and their candidate values have been set as follows: 
\begin{itemize}
\item {
Type of the RNN layer in the knowledge state encoder: LSTM, GRU.
}
\item {
Hidden state size $H_{\mathit{RNN}}$ of the chosen RNN layer: 32, 64, 128.
}
\item {
Hidden state size $H_{\mathit{MLP}}$ of the MLP layer: 32, 64, 128.
}
\item {
Dimensionality $d$ of the vector space in which the knowledge state and skills are embedded: 32, 64, 128.
}
\end{itemize}

The retention rate of 0.6 for the RNN dropout and the batch size of 128 were set to default. The Adam optimization method \cite{kingma2014adam} was used to minimize the total error $E_{total}$. 

Additionally, DKT was run with the skill vectors learned by KQN to evaluate their quality for the correctness prediction task. Specifically, at each time step, input $\mathbf{x}_t$ given to DKT was set to the concatenation of two vectors: one-hot encoded correctness $\mathbf{c}_t \in \mathbb{R}^N$ and the learned skill vector $\mathbf{s}_t \in \mathbb{R}^d$ corresponding to the original skill $e_t$. We denote DKT with such a setup as DKT+KQN. The hyperparameters of DKT+KQN have been searched in the same way as explained previously for KQN with the same dropout rate of RNN and the batch size. Meanwhile, LSTM was used for the RNN layer following past works \cite{piech2015deep,xiong2016going}.

\subsubsection{Knowledge Interaction Visualization}
Throughout a student's responses, prediction estimates for correctness with respect to the skills the student solved were calculated with the knowledge state query followed by the logistic function. A sample from the test set of ASSISTments 2009-2010 was used for the task. Then those estimates were visualized with a heat map to evaluate their changes as the student solved the problems.

\subsubsection{Skill Domain Analysis}
Skill distances have been used for clustering on the four datasets. To decide the linkage and the type of distance measures to use, we first performed flat clustering on Synthetic-5 with the number of clusters fixed to 5, the ground truth number of clusters. Then the quality of clustering or partitioning with respect to the original cluster labels was measured with the Adjusted Rand Index (ARI) \cite{hubert1985comparing}. It has a maximum value of 1 when the clusters are formed to match the original partitioning perfectly, and a minimum value of 0 when they are randomly partitioned. Since there are 20 sub-datasets for Synthetic-5, 20 ARI scores were averaged. The linkage between clusters and the type of distance measures have been set to hyperparameters as follows:
\begin{itemize}
\item {
Cluster linkage: $\{$average, centroid, complete, median, single, ward, weighted$\}$
}
\item {
Type of distance measure: $\{$cosine, Euclidean$\}$
}
\end{itemize}

After deciding which linkage and distance measure to use, the number of clusters $n$ has been explored. First, for different values of $n=5, \cdots,14$, the skills of ASSISTments 2009-2010 were clustered based on the distances computed from the skill vectors learned by KQN. Then DKT was used to quantify the quality of those clusters as follows: First, all the original skill IDs were substituted with the assigned cluster labels, where data splits remained the same as those for the correctness prediction task. Then, DKT was run five times. Finally, the average and the standard deviation of the test AUCs of DKT were reported.

\subsubsection{Sensitivity Analysis of the Vector Space Dimensionality}
Let $d$ be the dimensionality of the vector space in KQN and $d_{opt}$ be the optimal values of $d$ obtained in the correctness prediction task previously. KQN was trained on the four datasets by varying $d$ to $0.5d_{opt}$ and $2d_{opt}$ with the data splits kept the same, and the other hyperparameters set to the optimal values. To analyze the effect of $d$ on the correctness prediction task and the learning of skill vectors, prediction accuracy was reported, and the three distance matrices of the skill vectors for each of $d = d_{opt}, 0.5d_{opt}, 2d_{opt}$ were compared.

\section{Results and Analysis}
\subsection{Correctness Prediction}

\begin{table*}[tbp]
\centering
\resizebox{\textwidth}{!}{\begin{tabular}{|l|l|l|l|} 
\hline
2 Area Irregular Figure                                   & 41 Finding Percents                                & 74 Multiplication and Division Positive Decimals                           & 55 Divisibility Rules                             \\
46 Algebraic Solving                                      & 42 Pattern Finding                                 & 107 Parts of a Polynomial, Terms, Coefficient, Monomial, Exponent, Variable & 57 Perimeter of a Polygon                         \\ 
\cdashline{3-3}
56 Reading a Ruler or Scale                               & 43 Write Linear Equation from Situation            & 4 Table                                                                    & 71 Angles on Parallel Lines Cut by a Transversal  \\
63 Scale Factor                                           & 44 Square Root                                     & 12 Circle Graph                                                            & 72 Write Linear Equation from Ordered Pairs       \\
67 Percents                                               & 47 Percent Discount                                & 32 Box and Whisker                                                         & 80 Unit Conversion Within a System                \\
78 Rate                                                   & 54 Interior Angles Triangle                        & 49 Complementary and Supplementary Angles                                  & 83 Area Parallelogram                             \\
84 Effect of Changing Dimensions of a Shape Proportionally & 62 Ordering Real Numbers                           & 53 Interior Angles Figures with More than 3 Sides                          & 91 Polynomial Factors                             \\
85 Surface Area Cylinder                                  & 65 Scientific Notation                             & 58 Solving for a variable                                                  & 97 Choose an Equation from Given Information      \\
86 Volume Cylinder                                        & 76 Computation with Real Numbers                   & 59 Exponents                                                               & 101 Angles - Obtuse, Acute, and Right             \\
88 Solving Systems of Linear Equations                    & 79 Solving Inequalities                            & 68 Area Circle                                                             & 104 Simplifying Expressions positive exponents    \\ 
\cdashline{4-4}
92 Rotations                                              & 81 Area Rectangle                                  & 70 Equation Solving More Than Two Steps                                    & 20 Addition and Subtraction Integers              \\
93 Reflection                                             & 82 Area Triangle                                   & 75 Volume Sphere                                                           & 31 Circumference                                  \\
96 Interpreting Coordinate Graphs                         & 87 Greatest Common Factor                          & 102 Distributive Property                                                  & 34 Conversion of Fraction Decimals Percents       \\ 
\cdashline{1-1}\cdashline{3-3}
14 Proportion                                             & 89 Solving Systems of Linear Equations by Graphing & 1 Area Trapezoid                                                           & 60 Division Fractions                             \\
66 Write Linear Equation from Graph                       & 90 Multiplication Whole Numbers                    & 6 Stem and Leaf Plot                                                       & 77 Number Line                                    \\ 
\cdashline{1-1}\cdashline{4-4}
69 Least Common Multiple                                  & 98 Intercept                                       & 10 Venn Diagram                                                            & 19 Multiplication Fractions                       \\ 
\cdashline{1-1}
3 Probability of Two Distinct Events                      & 99 Linear Equations                                & 11 Histogram as Table or Graph                                             & 25 Subtraction Whole Numbers                      \\
5 Median                                                  & 100 Slope                                          & 21 Multiplication and Division Integers                                    & 61 Estimation                                     \\
7 Mode                                                    & 105 Finding Slope from Ordered Pairs               & 22 Addition Whole Numbers                                                  & 109 Finding Slope From Equation                   \\ 
\cdashline{4-4}
8 Mean                                                    & 106 Finding Slope From Situation                   & 26 Equation Solving Two or Fewer Steps                                     & 24 Addition and Subtraction Fractions             \\
9 Range                                                   & 108 Recognize Quadratic Pattern                    & 33 Ordering Integers                                                       & 50 Pythagorean Theorem                            \\
13 Equivalent Fractions                                   & 110 Quadratic Formula to Solve Quadratic Equation  & 37 Ordering Positive Decimals                                              & 103 Recognize Linear Pattern                      \\ 
\cdashline{2-2}\cdashline{4-4}
15 Fraction Of                                            & 16 Probability of a Single Event                   & 38 Rounding                                                                & 29 Counting Methods                               \\
18 Addition and Subtraction Positive Decimals             & 45 Algebraic Simplification                        & 39 Volume Rectangular Prism                                                & 95 Midpoint                                       \\ 
\cdashline{4-4}
23 Absolute Value                                         & 73 Prime Number                                    & 40 Order of Operations All                                                 & 64 Surface Area Rectangular Prism                 \\ 
\cdashline{4-4}
28 Calculations with Similar Figures                      & 94 Translations                                    & 48 Nets of 3D Figures                                                      & 36 Unit Rate                                      \\ 
\cdashline{2-2}\cline{4-4}
30 Ordering Fractions                                     & 17 Scatter Plot                                    & 51 D.4.8-understanding-concept-of-probabilities                            & \multicolumn{1}{l}{}                              \\
35 Percent Of                                             & 27 Order of Operations +,-,/,* () positive reals   & 52 Congruence                                                              & \multicolumn{1}{l}{}                              \\
\cline{1-3}
\end{tabular}}
\caption{14 flat clusters of ASSISTments 2009-2010 skills based on the average linkage method and the Euclidean distance. In each cluster, skills are sorted in an ascending order based on skill IDs. Different clusters are separated by dashed lines.}
\label{skill-clusters}
\end{table*}

Prediction accuracy was measured with the Area Under the ROC curve (AUC) during the testing phase. Note that the AUC of a model that guesses 0 or 1 randomly should be 50\%. As representatives of non-neural-network models, BKT, IRT, and their variants have been compared with KQN while DKT and DKVMN were compared to the state-of-the-art neural network models. The AUC results for those models have been cited from other papers as follows: those of IRT and its extensions from \cite{wilson2016back}, of BKT and its variants from \cite{khajah2016deep,xiong2016going}, and of DKT and DKVMN from \cite{zhang2017dynamic}.

Test AUCs for all the datasets are shown in Table \ref{prediction-performance}. Overall, KQN performed better than all the previously available KT models and showed a more stable performance with the lowest standard deviation values.

For ASSISTments 2009-2010, the test AUC of KQN was 82.32\% beating the previous highest value by 0.75\%. DKT+KQN showed the AUC of 82.05\%, not only higher than the original DKT performance but also higher than all the others. Surprisingly, for ASSISTments 2015, DKT+KQN achieved the highest test AUC of 73.41\%, even slightly higher than KQN. Both KQN and DKT+KQN performed better than all the previous results, which is promising in that KQN should be learning useful skill vectors that are transferable to other models and applications. For OLI Engineering Statics 2011, KQN achieved the highest value of 83.20\%, higher than the previous highest by 0.34\%. DKT+KQN showed a performance comparable to the vanilla DKT with the slightly higher test AUC of 80.27\% and the standard deviation of 0.22\%. Lastly, also for Synthetic-5, KQN performed the best with the highest average 82.81\% and the lowest standard deviation of 0.01\%. Interestingly, the standard deviation of DKT+KQN was much lower than that of the original DKT, showing that the learned skill vectors should be contributing to the stable prediction of the model.

In summary, KQN showed the best performance for the correctness prediction task compared to all the previous models while it achieved the second best result of all the models as DKT+KQN had the highest score on ASSISTments 2015. In addition to the best mean test AUC scores, our model had much lower standard deviation values compared to other models. DKT+KQN also had low standard deviation values for all the datasets except for OLI Engineering Statics 2011. Therefore, we speculate that KQN is able to produce stable prediction estimates due to its ability to learn a meaningful latent structure of the skill vectors.

\subsection{Knowledge Interaction Visualization}
\begin{figure}[ht]
\centering
\includegraphics[width=\linewidth, left]{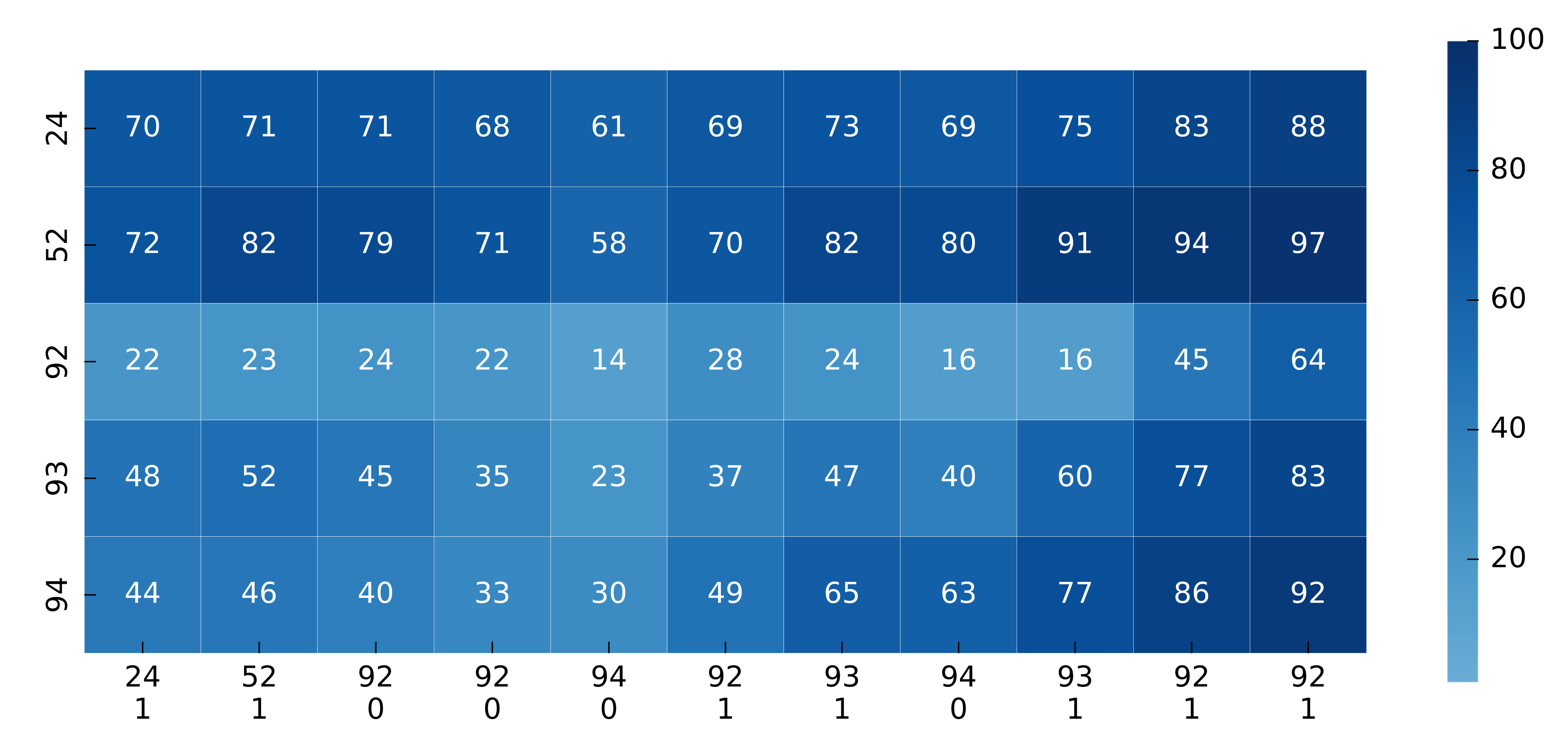}
\caption{Visualization of knowledge interaction by querying the knowledge state with respect to particular skills. On the x-axis, student responses are labeled while on the y-axis, all the skills contained in the responses are marked. Each column corresponds to one time step $t$, which increases along the x-axis. Prediction estimates for correctness in percentage (\%) are annotated in the grid. It is better viewed in color.}
\label{ks-query}
\end{figure}

\noindent For a sample student from ASSISTments 2009-2010, prediction estimates for correctness in percentage are visualized in Figure \ref{ks-query} through the knowledge state query with respect to particular skills. On the x-axis, student responses with skill IDs and correctness values as tuples are marked while on the y-axis, all the skills that the student solved are sorted in ascending order from the top. The corresponding skill names can be found in Table \ref{skill-clusters}.

Changes in probability estimates are mostly intuitive. For example, at $t=2$, after the student solved a problem with skill 52 correctly, the probability estimate for skill 52 increased from 72\% to 82\%. However, it can also be observed that some changes are counter-intuitive. For example, at $t=3$, as the student solved a problem with skill 92 incorrectly, the corresponding estimate increased from 23\% to 24\% even though the change was only 1\%. This problem has also been addressed for DKT in a previous work and is still an open problem to be improved \cite{yeung2018addressing}.

\begin{table}[tbp]
\centering
\resizebox{0.5\columnwidth}{!}{\begin{tabular}{lll} 
\hline
Linkage                            & Distance           & ARI              \\ 
\hline
\multirow{2}{*}{\textbf{average }} & cosine             & 0.3180           \\
                                   & \textbf{Euclidean} & \textbf{0.3266}  \\ 
\hline
\multirow{2}{*}{centroid}          & cosine             & 0.0373           \\
                                   & Euclidean          & 0.0143           \\ 
\hline
\multirow{2}{*}{complete}          & cosine             & 0.2898           \\
                                   & Euclidean          & 0.2898           \\ 
\hline
\multirow{2}{*}{median}            & cosine             & 0.0368           \\
                                   & Euclidean          & 0.0071           \\ 
\hline
\multirow{2}{*}{single}            & cosine             & 0.0703           \\
                                   & Euclidean          & 0.0703           \\ 
\hline
\multirow{2}{*}{ward}              & cosine             & 0.3201           \\
                                   & Euclidean          & 0.3234           \\ 
\hline
\multirow{2}{*}{weighted}          & cosine             & 0.2996           \\
                                   & Euclidean          & 0.3020           \\
\hline
\end{tabular}}
\caption{Average ARI scores for different linkage methods and distance measures on Synthetic-5.}
\label{adjusted-rand-score}
\end{table}

\begin{table}[ht]
\centering
\resizebox{0.6\columnwidth}{!}{\begin{tabular}{cc} 
\hline
Number of Clusters & Test AUC (\%)        \\ 
\hline
5           & 79.77$\pm$0.03           \\
6           & 79.97$\pm$0.03           \\
7           & 80.10$\pm$0.03           \\
8           & 80.04$\pm$0.02           \\
9           & 80.14$\pm$0.03           \\
10          & 80.10$\pm$0.02           \\
11          & 80.23$\pm$0.05           \\
12          & 80.20$\pm$0.05           \\
13          & 80.55$\pm$0.03           \\
\textbf{14} & \textbf{80.64$\pm$0.03} \\
\hline
\end{tabular}}
\caption{Test AUCs (\%) of DKT on ASSISTments 2009-2010 after replacing skill IDs with cluster labels assigned by flat clustering. The average linkage and the Euclidean distance were used.}
\label{test-auc-assist2009}
\end{table}

\subsection{Skill Domain Analysis}

\begin{table*}[tbp]
\centering
\begin{tabular}{llcccccc}
\multicolumn{1}{c}{Distance}  
& \multicolumn{1}{c}{Dataset} 
& $\xi_{d_{opt},0.5d_{opt}}$ 
& $\xi_{d_{opt},2d_{opt}}$ 
& $\xi_{0.5d_{opt},2d_{opt}}$        
& $\eta_{d_{opt}}$  
& $\eta_{0.5d_{opt}}$ 
& $\eta_{2d_{opt}}$ \\ 
\cline{3-8}
\hline
cosine    & ASSIST2009  & 0.11       & \textbf{0.09}     & 0.10        & 0.75 & 0.76   & 0.76  \\
          & ASSIST2015  & 0.11       & \textbf{0.09}     & 0.10        & 0.70 & 0.69   & 0.69  \\
          & Statics2011 & 0.15       & \textbf{0.09}     & 0.20        & 0.44 & 0.55   & 0.37  \\
          & Synthetic-5   & 0.12       & \textbf{0.10}     & 0.11        & 0.78 & 0.78   & 0.79  \\ 
\hline
Euclidean & ASSIST2009  & 0.09       & \textbf{0.07}     & 0.09        & 1.22 & 1.23   & 1.23  \\
          & ASSIST2015  & 0.09       & \textbf{0.07}     & 0.09        & 1.17 & 1.17   & 1.17  \\
          & Statics2011 & 0.16       & \textbf{0.11}     & 0.21        & 0.92 & 1.04   & 0.84  \\
          & Synthetic-5   & 0.10       & \textbf{0.08}     & 0.09        & 1.25 & 1.24   & 1.25               
\end{tabular}
\caption{Average pairwise distances and average differences between the pairwise distances.}
\label{pdists}
\end{table*}

Average ARI scores for clustering with different linkage methods and distance measures are reported in Table \ref{adjusted-rand-score}. ARI was the highest when the linkage was set to average, and the distance measure was set to Euclidean for clustering. Not surprisingly, the average ARI scores did not differ much when the distance measure was set to either cosine or Euclidean.

After clustering the skills of ASSISTments 2009-2010 with the linkage and the distance measure set to average and Euclidean, respectively, and substituting the original skill IDs with those cluster labels, DKT was run five times. The test AUCs of DKT are reported in Table \ref{test-auc-assist2009}. They increased gradually as the number of clusters changed from 5 to 14. The lowest test AUC was 79.77\% when $n=5$, not differing much from the highest test AUC of 80.64\%, which means that skills were clustered preserving useful information.

On ASSISTments 2009-2010, skill IDs and skill names were grouped by 14 clusters as shown in Table \ref{skill-clusters}. Different clusters were separated by dashed lines while in each cluster, skills were sorted in ascending order based on their skill IDs. Since the skills must have been grouped based on probabilistic skill similarity, a number of intuitively similar skills were clustered together. For example, `30 Ordering Functions' and `62 Ordering Real Numbers' were assigned to the fourth cluster while `33 Ordering Integers' and `37 Ordering Positive Decimals' were assigned to the eighth cluster.

\subsection{Sensitivity Analysis of the Vector Space Dimensionality}
\noindent For the four datasets, the test AUCs of KQNs with the embedding dimensionality $d=d_{opt}, 0.5d_{opt},$ and $2d_{opt}$ are shown in Table \ref{KQN-performance}, where $d_{opt}$ refers to the optimal values chosen from the holdout validation for the correctness prediction task. We could observe only a little difference in the prediction accuracy as the values of $d$ were varied.

For each pair of $\{d_{opt}, 0.5d_{opt}, 2d_{opt}\}$, the average difference $\xi$ between the pairwise distances of the skill vectors of two different dimensionalities was calculated as follows:
\begin{align*}
&^\forall d_1, d_2 \in \{d_{opt}, 0.5d_{opt}, 2d_{opt}\}, \text{ } d_1 \neq d_2, \\
&\xi_{d_1, d_2} = \frac{\sum_{i > j} |\mathit{pdist}_{d_1}(\mathbf{s}_i, \mathbf{s}_j) - \mathit{pdist}_{d_2}(\mathbf{s}_i, \mathbf{s}_j)|}{\binom{N}{2}},
\end{align*}
where $\mathit{pdist}_d(\mathbf{s}_i, \mathbf{s}_j)$ refers to the pairwise distance between two skill vectors $\mathbf{s}_i$ and $\mathbf{s}_j$, and $N$ is the number of skills. $\xi$ is then compared to the average pairwise distance $\eta$ as follows:
\begin{align*}
\eta_d = \frac{\sum_{i > j}^N \mathit{pdist}_d(\mathbf{s}_i, \mathbf{s}_j)}{\binom{N}{2}}.
\end{align*}

In Table \ref{pdists}, the lowest values of $\xi$ are indicated in bold. As can be seen, $\xi_{d_{opt}, 2d_{opt}}$ is always lower than $\xi_{d_{opt}, 0.5d_{opt}}$. From this, it can be inferred that KQN learned the skill relationships better when $d$ was set to a number high enough since $d$ controls the maximum number of mutually independent skill vectors. Also, the values of $\xi$ were relatively low compared to the corresponding values of $\eta$. For example, $\xi_{d_{opt}, 2d_{opt}}$ was only $0.07$ when the Euclidean distance was used for ASSISTments 2009-2010 while the corresponding $\eta_{d_{opt}}$ and $\eta_{2d_{opt}}$ were 1.22 and 1.23, respectively.

To further evaluate the distance matrices, we performed Mantel tests \cite{legendre1998numerical}, which measure the similarity between two distance matrices with a correlation coefficient $\rho$ and a p-value. $\rho$ has the same range as that of correlation coefficients in statistics while a p-value indicates statistical significance. The Pearson correlation and the permutation number of 999 were set for the Mantel tests.

The results of the Mantel tests are reported in Table \ref{mantel}, where p-values are omitted since they were $0.001$ in all cases, indicating that the values of $\rho$ are statistically significant. The fact that the values of $\rho$ were always the highest for $d_{opt}$ and $2d_{opt}$ confirmed that there was the strongest positive correlation between the corresponding distance matrices. Specifically, $\rho$ was over $0.8$ for OLI Engineering Statics 2011 while it had the minimum value of $0.536$ for all the other datasets, proving strong positive correlations as well. Therefore, from Table \ref{pdists}, Table \ref{KQN-performance}, and Table \ref{mantel}, KQN was shown to be stable in predicting correctness and learning the relationships between the skill vectors as the value of the vector space dimensionality $d$ was varied.

\begin{table}[ht]
\centering
\begin{tabular}{lccc} 
\hline
\multicolumn{1}{c}{\multirow{2}{*}{Dataset}} & \multicolumn{3}{c}{Test AUC (\%)}  \\ 
\cline{2-4}
                         & \multicolumn{1}{c}{$d_{opt}$} & \multicolumn{1}{c}{$0.5d_{opt}$} & \multicolumn{1}{c}{$2d_{opt}$}       \\ 
\hline
ASSIST2009               & 82.32  & 82.35     & 82.32         \\ 
\hline
ASSIST2015               & 73.40  & 73.38     & 73.40         \\ 
\hline
Statics2011              & 83.20  & 83.17     & 83.16         \\ 
\hline
Synthetic-5              & 82.81  & 82.79     & 82.82         \\
\hline
\end{tabular}
\caption{Test AUCs of KQNs by varying the embedding dimensionality $d$. $d_{opt}$ refers to the optimal values found from the correctness prediction task. Note that prediction accuracy may not be the highest when $d$ was set to $d_{opt}$.}
\label{KQN-performance}
\end{table}

\section{Conclusions and Future Work}
\begin{table}[ht]
\centering
\resizebox{\columnwidth}{!}{\begin{tabular}{llccc}
\multicolumn{1}{c}{\multirow{2}{*}{Distance}} 
& \multicolumn{1}{c}{\multirow{2}{*}{Dataset}}
& \multicolumn{3}{c}{$\rho$}   \\ 
\cline{3-5}
&    & $d_{opt}$-$0.5d_{opt}$   & $d_{opt}$-$2d_{opt}$ & $0.5d_{opt}$-$2d_{opt}$   \\ 
\hline
cosine                    & ASSIST2009               & 0.521  & \textbf{0.653} & 0.570  \\
                          & ASSIST2015                  & 0.582  & \textbf{0.661} & 0.609 \\
                          & Statics2011                 & 0.616  & \textbf{0.825} & 0.609 \\
                          & Synthetic-5                   & 0.495  & \textbf{0.526} & 0.511 \\ 
\hline
Euclidean                 & ASSIST2009                  & 0.531  & \textbf{0.660} & 0.583  \\
                          & ASSIST2015                  & 0.601  & \textbf{0.682} & 0.629  \\
                          & Statics2011                 & 0.620  & \textbf{0.816} & 0.607  \\
                          & Synthetic-5                   & 0.508  & \textbf{0.536} & 0.523
\end{tabular}}
\caption{Mantel tests on the distance matrices. p-values are not marked since they were 0.001 in all cases.}
\label{mantel}
\end{table}

From the experiment results for the four tasks, we list the contributions of this paper as follows:
\begin{enumerate}
\item {
KQN performs better than all the previous models on the four datasets for the correctness prediction task.
}
\item{
KQN enables the knowledge state of a student to be queried with respect to different skills, which is helpful for interpreting the knowledge interaction through visualization.
}
\item{
KQN's architecture leads to the concept of \textit{probabilistic skill similarity} to relate the cosine and Euclidean distances between two skill vectors to the odds ratio for the corresponding skills as introduced previously in the paper. This makes the skill vectors and their pairwise distances useful for domain modeling, e.g., with cluster analysis.
}
\item{
KQN is robust to the changes in the dimensionality of the vector space for the knowledge state and skill vectors in that its prediction accuracy is not degraded and it learns strongly positively correlated sets of pairwise distances between the skill vectors as the value of the dimensionality is varied, or equivalently, KQN learns the latent relationships between skills stably.
}
\end{enumerate}

Compared to other neural network models, KQN has more parameters to learn. For example, since it includes an MLP in the skill encoder in addition to an RNN in the knowledge state encoder, KQN is computationally heavier than DKT which only has an RNN for encoding student responses. Heuristically, more GPU memory was required for training KQN compared to DKT+KQN. Still, we believe that the advantages of KQN mentioned above are meaningful enough to compensate for the increase in space complexity.

KQN proposes an alternative approach to the KT problem by defining the knowledge state and skill vectors in the same vector space. It has a general form of the knowledge state and skills as vectors while defining the knowledge interaction clearly as the dot product between the two types of vectors. From the fact that the pairwise distances between skill vectors are interpreted as the logarithm of the corresponding odds ratios from the probabilistic perspective, those distances can become useful features for domain modeling to explore the latent structure of the skill domain, which can be a future direction of the KT research.

\section{Acknowledgements}
This research has been supported by the project ITS/227/17FP from the Innovation and Technology Fund of Hong Kong.

\bibliographystyle{ACM-Reference-Format}
\bibliography{bib}

\end{document}